%% file: main_final.tex
\documentclass[conference]{IEEEtran}
\usepackage{amsthm,amssymb,graphicx,multirow,amsmath,color,amsfonts,subfig,balance}
\usepackage[latin1]{inputenc}
\usepackage[noadjust]{cite}
\usepackage{tikz}
\usetikzlibrary{arrows,calc}		% Optional ticks libraries
\usepackage{algorithm} 
\usepackage{algorithmic}  
\usepackage[algo2e]{algorithm2e}

\usepackage{bbm} % for \mathbbm{1}
\usepackage{pdfpages}
\usepackage{tabulary}
\usepackage{multirow}
\usepackage{comment}
\usepackage{todonotes}
\def\beq{\begin{equation}}
\def\eeq{\end{equation}}
\def\beqa{\begin{eqnarray}}
\def\eeqa{\end{eqnarray}}
\def\beqan{\begin{eqnarray*}}
\def\eeqan{\end{eqnarray*}}
\include{notation_new}

\begin{document}
\title{Power Efficient Discontinuous Reception 
in THz and mmWave Wireless Systems}
\author{Syed Hashim Ali Shah, Sundar Aditya, Sourjya Dutta, Christopher Slezak and Sundeep Rangan \\
NYU WIRELESS, Tandon School of Engineering, New York University, Brooklyn, NY 11201\\
{\footnotesize e-mail: \{s.hashim, sundar.aditya, sdutta, chris.slezak, srangan\}@nyu.edu}
}
\maketitle

\begin{abstract}
Discontinuous reception (DRX), where a 
user equipment (UE) temporarily disables its receiver, is a 
critical power saving feature in modern cellular systems.
DRX is likely to be particularly aggressively used in 
the mmWave and THz frequencies due 
to the high front end power consumption.
A key challenge of DRX in these frequencies
is that individual links are directional and
highly susceptible to blockage.
MmWave and THz UEs will therefore likely
need to monitor multiple cells in multiple directions
to ensure continuous reliable connectivity.
This work proposes a novel, heuristic
algorithm to dynamically select the
cells to monitor to attempt to optimally trade-off 
link reliability and power consumption.
The paper provides preliminary estimates of 
connected mode DRX mode consumption using detailed
and realistic statistical models of blockers
at both 28 and 140~GHz.
It is found that although blockage dynamics are faster
at 140~GHz, reliable connectivity at low power can be maintained with sufficient macro-diversity and link prediction.
\end{abstract}

\begin{IEEEkeywords}
Discontinuous reception (DRX), terahertz (THz) communications, millimeter wave (mmWave) communications 
\end{IEEEkeywords}

\section{Introduction}
\label{sec:motivation}
Mobile wireless communication in the
mmWave and THz bands enables multi-Gbps 
peak throughput, but at the cost 
of high power consumption
in both radio frequency front-end (RFFE)
and digital baseband processor.
As we will see below, 
peak power consumption
in the UE in the THz frequencies
can exceed 1 Watt, a large portion of
the total smartphone power budget.
Discontinuous reception (DRX) modes \cite{drx1,drx2}
where a mobile device or user equipment (UE) temporarily disables its receiver radio frequency front end (RFFE), 
can offer significant power savings in 
cases where the traffic is intermittent.

A key challenge in implementing DRX
in the mmWave and THz frequencies is that
links are highly susceptible to blockage
by many common materials, as well as the human
body or hand \cite{rappaport2014millimeter}.
Thus, small changes in the environment
or orientation of the handset can lead
to rapid drops in link quality
\cite{slezak2018empirical,raghavan2018spatio,maccartney2017rapid}.  
Hence, in DRX mode, the UE will likely need
to monitor multiple cells to maintain
reliable connectivity. Indeed,
macro-diversity has been long identified 
as key in mmWave cellular systems
\cite{choi2014macro}.
However, from an energy standpoint,
monitoring
multiple links reduces the inactive time,
thus creating a trade-off of power consumption
and reliability.

\begin{figure}
 \centering
    %\begin{subfigure}{0.4\textwidth}
     %\centering
       \includegraphics[scale=0.5]{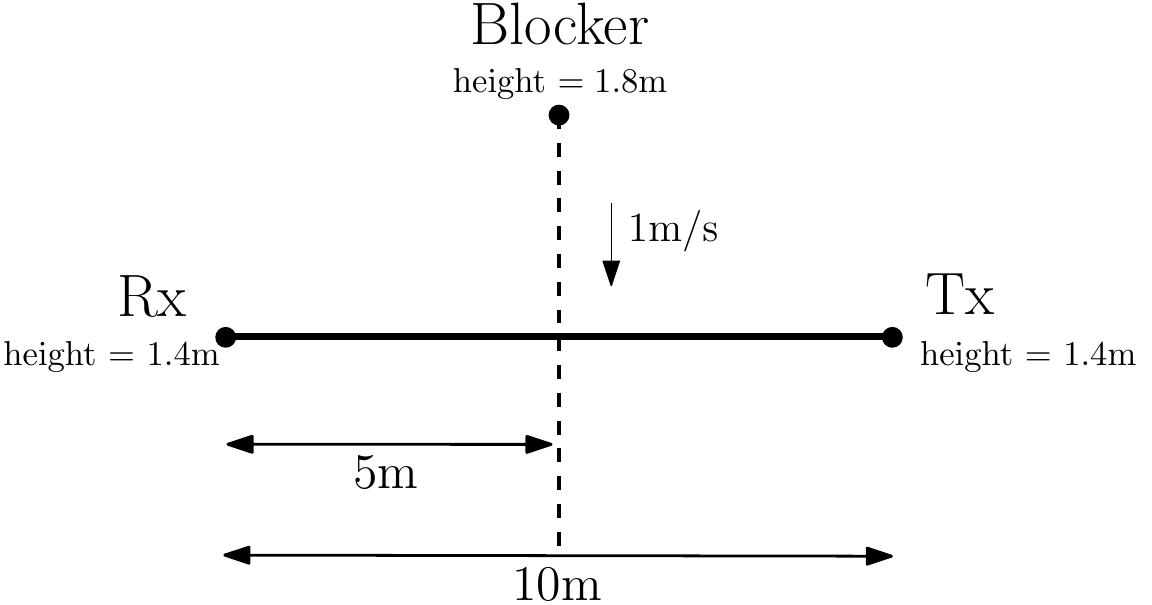}
    %\caption{Movement of a human blocker, with dimensions in accordance with \cite{3gpp_model}.}
    %\label{fig:human_movement}
    %\end{subfigure}
    %\begin{subfigure}{0.4\textwidth}
    %\centering
        \includegraphics[trim={3.5cm 0 3cm 0cm},clip,width=0.4\textwidth]{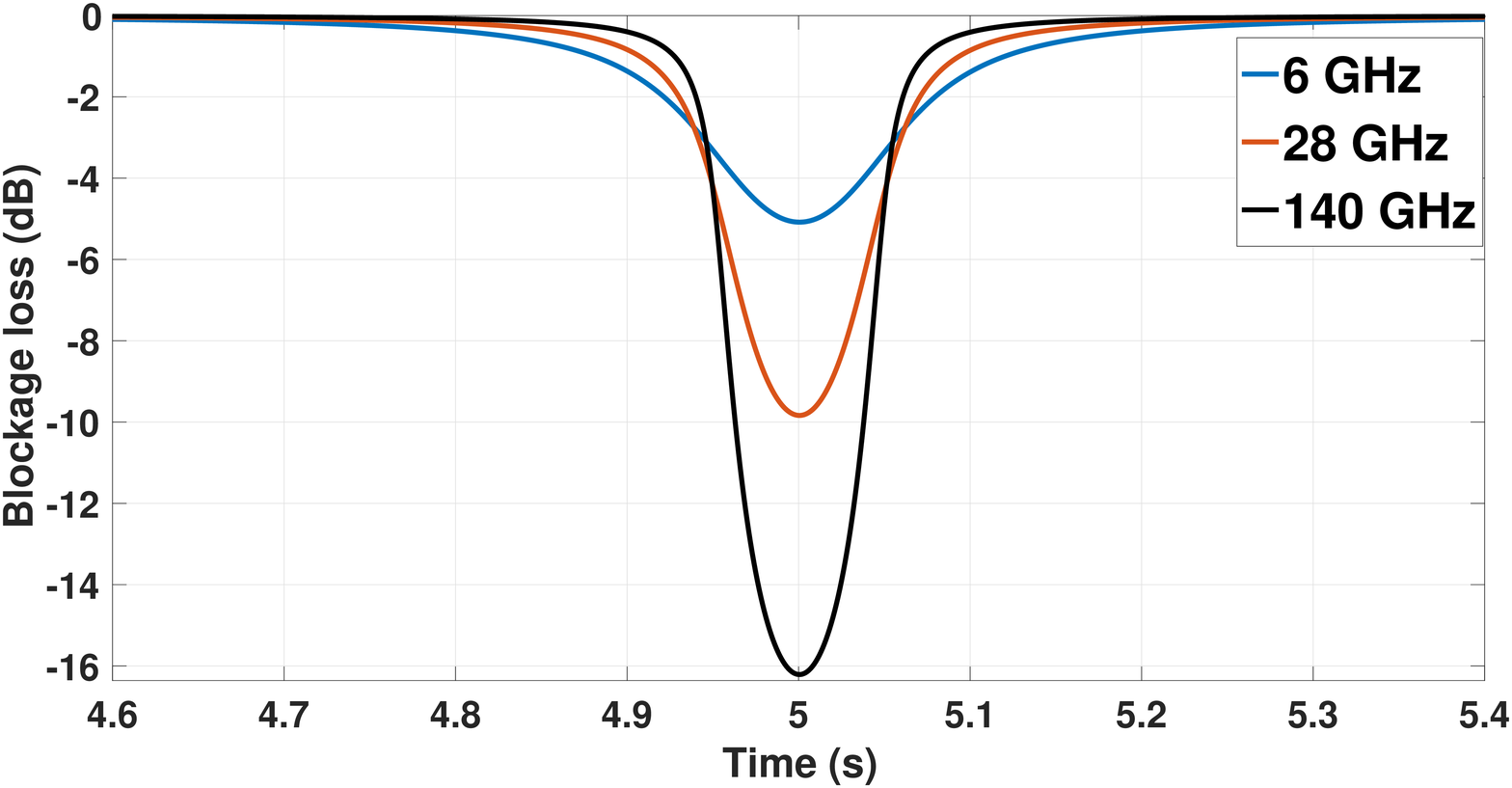}
    %\caption{Blockage loss due to Fig.~\ref{fig:human_movement}.}
    %\label{fig:blockage_fade}
%\end{subfigure}
\caption{Understanding the severity of blockages at higher frequencies.  Top panel:  Movement
of a human blocker following the 3GPP model \cite{3gpp_model}.  Bottom:  Blockage loss
due to knife-edge diffraction.}
\label{fig:blockage}
\end{figure}

As communications move from the mmWave 
to the sub-THz bands,
the channel dynamics will likely become much faster.
For example, Fig.~\ref{fig:blockage}
shows the predicted fade with a human blocker
following the 3GPP double knife-edge diffraction
model~\cite{3gpp_model}.  We see that 
at 140~GHz, the blockage is both deeper and faster
than 28~GHz.  Faster channel dynamics 
will be
harder to track and necessitate monitoring more
cells for a given reliability.
In addition, with the larger number of antenna
elements and wider bandwidths, the RFFE
power consumption will be higher.
Hence, finding power efficient methods for properly
selecting cells to monitor in DRX mode, will be even more vital.

% Include blockage loss vs frequency figure here

%The increase in energy consumption due to beam search is exacerbated by the increase in the power consumption of the RFFE at higher frequencies, due to a combination of (i) requiring a larger number of antenna elements to achieve highly directional beams, leading to an increase in the number of low-noise amplifiers (LNAs) and phase shifters at the RFFE, and (ii) requiring analog-to-digital converters (ADCs) to operate at higher sampling frequencies due to the larger available bandwidth. For instance, in Section~\ref{sec:power_numbers}, we show that the RFFE power consumption increases nearly four-fold as we move from 28 to 140 GHz. Hence, from an energy consumption perspective, smarter policies are needed whereby a UE identifies a strong BS without having to listen to all BSs.

%%%%%%%%%%%%%%%%%%%%%%%%%%%%%%%%%%%%%%%%%%%%%%%%%%%%%%%%%%%%%%
%\item Emphasize the importance of power saving and how it will be important in future %once digital beamforming architectures are deployed \cite{powerconsumed}
 
%%%%%%%%%%%%%%%%%%%%%%%%%%%%%%%%%%%%%%%%%%%%%%%%%%%%%%%%%%%%%% 
% \item Give an overview of what is coming in the next sections of the paper 

The contribution of this paper is threefold.
First, we provide a preliminary estimate
of the power consumption per unit time
for DRX mode measurements.  We provide
estimates at both 28 and 140~GHz.
Second, to minimize the measurement wake
time,
we propose a simple algorithm where the UE tries to maintain its association to its current serving BS, while also tracking a subset (of a given cardinality) of the remaining BSs to mitigate the effects of blockage effects and also save power.
Third, we provide detailed simulations
of the algorithm using 3GPP path loss
and blockage models \cite{3gpp_model}.
%Through simulations of a single UE in a multi-cellular environment with random blockers, we observe that it is sufficient to track only four BSs to ensure that the link loss probability is at most $1\%$. 

%The rest of the paper is organized as follows. In Sec.~\ref{sec:power_numbers}, we compare the power consumed by multi-antenna front ends at mmWave and THz frequencies. The system model with the problem formulation is presented in Sec.~\ref{sec:problem_formulation}. We present the proposed algorithm for listening to BSs at DRX cycles in 
%Sec.~\ref{sec:Algo}. Simulation setup and results are presented in Sec.~\ref{sec:numerical_results}, 
%Finally, Sec.~\ref{sec:concl} concludes the paper with a few remarks on system level design implications stemming from our results.

\section{Front End Power Consumption at THz}
\label{sec:power_numbers}
\begin{figure}
    \centering
   \includegraphics[scale=0.5]{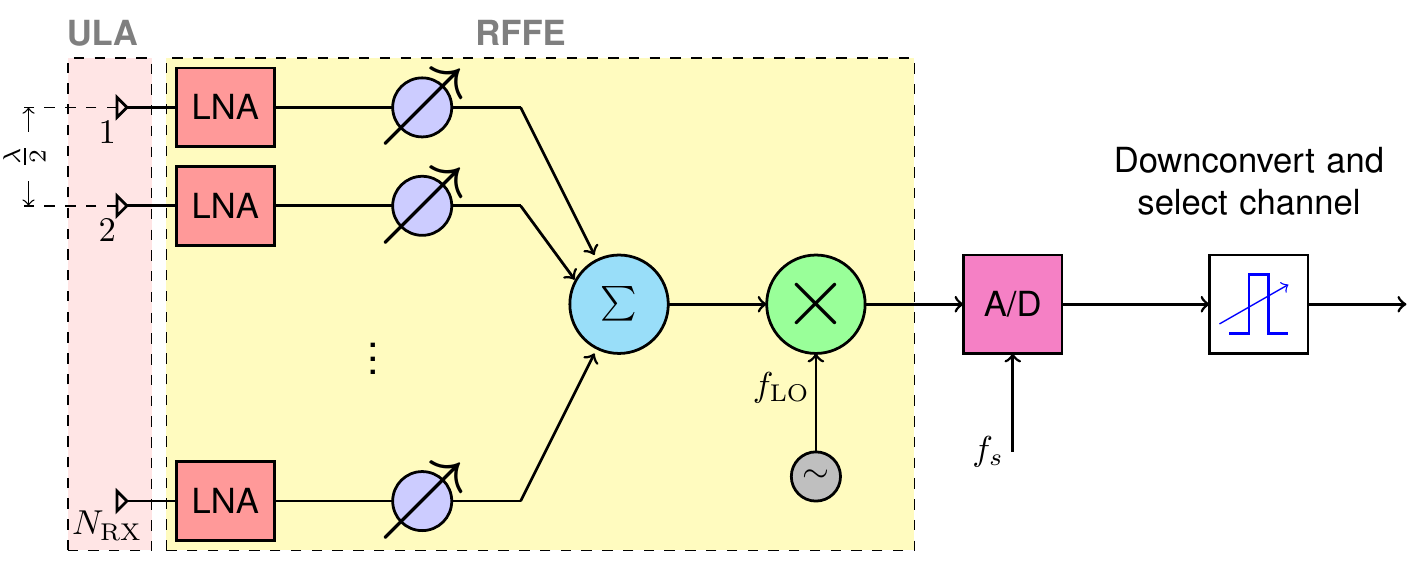}
    \caption{Analog beamforming based multi-channel receiver front end with channel selection performed in digital base band.}
    \label{fig:thzfe}
\end{figure}

We first attempt to estimate
the power consumption for link monitoring
in both mmWave and THz frequencies.
For ultra-wideband systems, receivers (RXs) should be designed such that they can be programmed to tune into any assigned channel within the available frequency band. One such  architecture is proposed in \cite{HwuRazavi2015}, where down conversion and channel selection are performed by the baseband digital circuit. We will focus on
\emph{analog beamforming} as shown in 
Fig.~\ref{fig:thzfe}, although similar calculations
can be done for digital beamforming as well.
For multi-channel operation, we assume
the RFFE and the ADCs operate over the entire band.

\paragraph*{ADC} The power consumed by an ADC scales linearly with the sampling frequency ($f_s$). For mmWave communications, $2$~GHz of bandwidth is allocated around the center frequency of $28$~GHz. Hence, ADCs for mmWave RFFEs need to operate at $f_s = 2$~GHz. In contrast, $7$~GHz of unlicensed spectrum is available in the THz band between $141$ and $148$~GHz \cite{spectrumallocation_2018}. 
Note that for the direct conversion architecture considered here, the ADCs operate over the entire available bandwidth although the individual component carriers may have a smaller bandwidths.
This leads to a $3.5\times$ increase in power consumption, assuming that ADCs with the same figures of merit (FoMs) and resolution are used at both bands. In Table \ref{tab:pow}, we consider an ADC FoM of $65$~fJ/steps and we report the power consumed by a pair of 8-bit ADCs operating at $f_s = 2$ and $f_s=7$~Gs/s at the Rx.

\paragraph*{RFFE} For a given distance, the free space path loss encountered at $140$~GHz is nearly $14$~dB higher compared to $28$~GHz mmWave bands. To mitigate this loss in link budget through beamforming, THz transmitters and receivers will require at least a quadrupling of the number of antenna elements on both sides
of the link. However,
due to the decreased wavelength,
the total antenna aperture can be decreased
or even reduced if there is gain
increase on both sides.  
For example, a $4 \times 4$ 
uniform plane array (UPA) with $\lambda/2$
spacing at 28~GHz would require approximately an
$2\times 2$ cm$^2$ area, while an
$8 \times 8$ 
array at 140~GHz would require approximately an
$0.8 \times 0.8$ cm$^2$ area.
%Thus, to maintain the same link budget as a $8$-element UE and $64$-element BS at $28$~GHz, a UE at $140$~GHz will require $N_{\rm Rx}= 64$ antennas when listening to a BS with a $256$ element array. It is interesting to note that due to the decrease in wavelength at $140$~GHz, a $\lambda/2$ spaced $64$ element uniform linear array (ULA) will be less than $7$~cm ($2.75$ inches) in length. 
%As a reference point, note that a $8$ element $\lambda/2$ ULA at 28 GHz is $8.5$~cm long.

An increase in the number of antennas for an analog beamformer implies an increase in the number of low noise amplifiers (LNAs) and RF phase shifters (PSs). Following the work in \cite{dutta2019}, we assume that the PSs, combiners and mixers are passive circuits. We also make the assumption that the FoM for the LNAs and the insertion loss (IL) due to the PS-s and mixers are the same as at 28 GHz \cite{dutta2019} and $140$~GHz. In Table \ref{tab:pow}, we compare the power drawn by the RFFE of a Rx at 28 GHz with $8$-antennas and one at $140$~GHz with $64$ antennas assuming a $10$~dBm of local oscillator power draw.

\begin{figure}[!t]
\centering
\includegraphics[scale=0.5,width=0.4\textwidth]{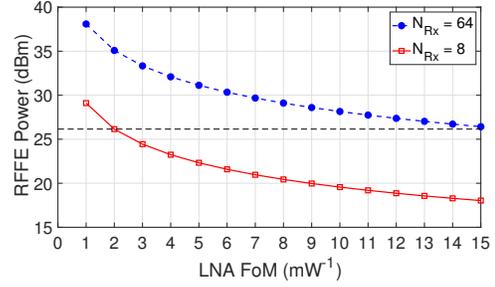} 
\caption{RFFE power consumption as a function of the LNA FoM. 
\\
For mathematical expressions see \cite{dutta2019}}
\label{fig:lna}
\end{figure}

%can remove this para and figures if not required.
One might argue that the RFFE power draw may be reduced by sophistication in circuit design that enhances the performance of the LNAs and the PSs. But this may not be the case. As shown in Fig.~\ref{fig:lna}, a 64 element system with a very high LNA FoM of $15~\text{mW}^{-1}$ draws the same power as an $8$ element system with a low LNA FoM of $2~\text{mW}^{-1}$. Similar observations can be made about PS IL as well. In fact, even with considerable advancements in devices and circuit design, a $64$ element RFFE will draw considerably more power than a $8$ element one. 

\begin{table}[!t]
    \centering
    \begin{tabular}{|| p{1.5cm}| p{1.5cm}| p{1.5cm}||} \hline
    Component      & $28$ GHz ($N_{\rm rx} =8$)& $140$ GHz ($N_{\rm rx} =64$)  \\ \hline \hline
    RFFE  &  133.7   & 999.3  \\ \hline
    ADC   &   66.6  &  232.9 \\  \hline
    Total &   200.3 & 1232.2 \\ \hline
    \end{tabular}
    \caption{Power consumption by Rx front end at mmWave and THz (all units in mW)}
    \label{tab:pow}
\end{table}

Due to the use of a large number of antennas and very wideband data converters, cellular front ends at $140$~GHz can consume nearly a Watt of additional power compared to those at $28$~GHz as evident from Table \ref{tab:pow}. Hence, power saving at the receiver by optimizing the discontinuous reception (DRX) procedure can be crucial for such systems, especially when employed in handheld UEs.

\section{System Model}
\label{sec:problem_formulation}

%\input{problem_formulation}
%For concreteness, we will assume a 5G-like protocol, where the UE is in RRC Idle and registered with the BSs in some tracking area. However, much of the theory we discuss here could also be applied in DRX connected mode when the UE is \emph{listening} to multiple BSs.
DRX is used at the UE both in the RRC connected mode (i.e., between active data transmissions) as well as the RRC idle and RRC inactive modes 
(i.e., when there is a long period of inactivity) \cite{3GPP38300}. In this work, we focus on the connected mode DRX, which is key for UE power savings especially under practical bursty traffic considerations and multi-user scenarios.
To support macro-diversity resistance 
against blockage, we assume
the network maintains multiple simultaneous connections
to the UE from different cells
(e.g. via carrier aggregation).  The set of cells
from which data can be transmitted is called
the \emph{serving set}.

%After transmitting or receiving all pending packets, the UE listens on the downlink (DL) channel for further DL grants. If no DL grant is received within a certain observation period, the UE goes into the DRX mode.

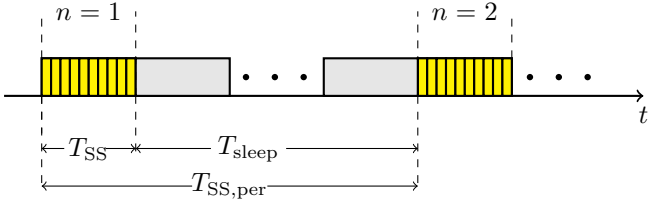
\begin{figure}[!t]
    \centering
    \begin{tikzpicture}[scale=0.5]
    \draw[thick,->] (0,0) -- (17,0) node [align=center] at (17,-0.5){$t$};
    
    \draw[thick,fill=gray!20](1,0) rectangle (3.5,1);
    \foreach \i in {1,1.25,1.5,1.75,2,2.25,2.5, 2.75, 3,3.25}
        \draw[thick,fill=yellow] (\i,0) rectangle (\i+0.25,1);
    
    \draw[thick,fill=gray!20](11,0) rectangle (13.5,1);  
    \foreach \i in {1,1.25,1.5,1.75,2,2.25,2.5, 2.75, 3,3.25}
        \draw[thick,fill=yellow] (10+\i, 0) rectangle (10.25+\i, 1);
    
    \node [align=center] at (7.2,0.5) {\Huge$\ldots$};
    \node [align=center] at (14.75,0.5) {\Huge$\ldots$};
    
    \foreach \j in {3.5, 8.5}    
        \draw[thick,fill=gray!20](\j,0) rectangle (\j+2.5,1);
        
    \draw[dashed] (1,0) -- (1,-2.4);  
    \draw[->] (1.6,-1.4) -- (1,-1.4); \draw[->] (2.7,-1.4) -- (3.5,-1.4) node  [align=center] at (2.25,-1.4){$T_{\rm SS}$};
    
   \draw[dashed] (1,1) -- (1,2.25);   %\draw[dashed] (2.75,1) -- (2.75,1.9) node [align=center] at (2,1.5){$T_{\rm SS}$};  
    \draw[dashed] (11,0) -- (11,-2.4);
    \draw[->] (5.5,-1.4) -- (3.5,-1.4);  \draw[->] (7.5,-1.4) -- (11,-1.4) node [align=center] at (6.5,-1.4) {$T_{\rm sleep}$}; 
    
    \draw[->] (5,-2.4) -- (1,-2.4);  \draw[->] (7,-2.4) -- (11,-2.4) node [align=center] at (6,-2.4) {$T_{\rm SS,per}$}; 
    
    \draw[dashed] (1,0) -- (1,-1); 
    \draw[dashed] (3.5,0) -- (3.5,-1.5);
    
     \draw[dashed] (3.5,0) -- (3.5,2.25) node [align=center,text width=1cm] at (2.25, 2.25) {$n = 1$};
     
     \draw[dashed] (11,0) -- (11,2.25) node [align=center,text width=1cm] at (12.25, 2.25) {$n = 2$};
     \draw[dashed] (13.5,1) -- (13.5,2.25);
        
    \end{tikzpicture}
    %\caption{Illustration of the $n$-th paging cycle, where SS,$k$ (PG,$k$) denotes the time at which the synchronization (paging) signal from BS $k$ is received at the UE. Without loss of generality, we assume that the $k$-th BS is the $k$-th in line to transmit its synchronization and paging signals.}
    \caption{Time line of a UE in connected mode DRX.} 
    %Here, $T_{\rm f}$ denotes a frame duration, $T_{\rm SS}$ is the duration of a SSB (yellow). The idle period $T_{\rm idle}$ spans multiple frames.
    \label{fig:paging_cycle}
\end{figure}

During DRX, the UE turns off its Rx FE and goes into the ``sleep mode" to save power. Periodically, the UE ``wakes up'' at pre-allocated time instances to either monitor a subset of the available links, 
or transmit and receive control signals.
During the monitoring intervals, the UE measures the links over an interval of duration $T_{\rm SS}$ as shown in Fig.~\ref{fig:paging_cycle}.
We assume the UE will measure link quality
from the synchronization signal bursts (SSBs)
transmitted by the tracked BS. %Assume synchronous transmission of SSBs by the BSs. 
The typical periodicity for the SSBs is
$T_{\rm SS,per}=20~{\rm ms}$ \cite{3gpp_RRC, bmanagement}.

We assume each BS has a different pointing angle to the UE.
Therefore the number of scans the UE needs to performs increase with the number of cells to monitor. 
To save power, we assume that the UE 
can monitor a a subset of the serving cells, 
which we call the \emph{listening set}.
Hence, the value of $T_{\rm SS}$ will be
determined by the number of cells in the listening set,
and  the UE will be in sleep mode for a duration $T_{\rm sleep}=T_{\rm SS,per}-T_{\rm SS}$ 
in each DRX \emph{cycle}. 
%At each measurement period in a DRX cycle, the UE checks if it has a usable or unblocked link. If the UE determines that it has lost its radio link, the UE has to do an \emph{exhaustive beam sweep} to find the best direction of communication.
 %\textcolor{red}{we will assume that
%it declares radio link failure and 
%performs initial access for connection
%re-establishment}\textcolor{blue}{(do these terms carry the same meaning in standards, also should we change the way we talk about initial access)}.

%and monitors the link with its associated as well as neighboring BSs and listens for DL messages. For the rest of the time,  

%The interval between successive waking up periods is referred to as a DRX cycle (Fig.~\ref{fig:paging_cycle}).

%\paragraph*{Beam Sweeping} For beamformed systems, beam sweeping can be very expensive both in terms of latency and power consumption. This is due to the large number of narrow beams that needs to be tracked. More so, for analog beamforming based architectures, an exhaustive beam search can take hundreds of milliseconds \cite{bmanagement,barati_initialaccess}. Fully digital architectures on the other had can have faster beam sweeping mechanism but draw similar power as exhaustive beam sweep using analog beamforming.\cite{barati_initialaccess}.

\paragraph*{Blockage}
Our goal is to quantify the trade-off of 
monitor power consumption and blockage.
To model the blockage,
let $M_{\rm u}$ and $M_{\rm b}$ denote the number of antenna elements at the UE and each cell or BS, respectively.
Within the DRX cycle, we index the monitoring or ``awake'' periods by $n = 1, 2, \ldots$  as shown in Fig.~\ref{fig:paging_cycle}. We term them as ``monitoring instances''. Let $\nbH^{(n,k)} \in \nbbC^{M_{\rm u} \times M_{\rm b}}$ denote the channel matrix between the UE and the $k$-th cell
in the $n$-th monitoring instance. The signal received at the UE is given by
\beq
\label{eq:y}
 y_{nk}=\nbw_k^H \nbH^{(n,k)} \nbf_k x + \xi_{nk},
\eeq
where  $\nbw_k \in \nbbC^{M_{\rm u} \times 1}$ and $\nbf_k \in \nbbC^{M_{\rm b} \times 1}$ are the beamforming vectors applied at the UE and the cell, respectively, for the given link, $x$ is the unit energy measurement signal, and $\xi_{nk} \sim \mathcal{CN}(0,\sigma^2)$ is the system noise. Hence, for the measurement signal transmitted from the $k$-th cell, the signal-to-noise ratio (SNR) at the UE is
\beq
\label{eq:snr}
 \gamma_{nk} = \frac{|\nbw_k^H \nbH^{(n,k)} \nbf_k|^2}{\sigma^2}.
\eeq

%During each monitoring instance, a UE chooses to monitor a set of $K$ neighboring cells $(1\leq K \leq N)$. 
Let $\mathcal {A}_n \subseteq \{1,\ldots,N\}$ denote the listening set, which is the 
set of cells chosen in the $n$-th monitoring instance. We refer to ${\mathcal A}_n$ as the  \emph{listening set}. We let $K$ denote the number of cells in
the listening set.  
Listening to fewer cells during each monitoring instance implies that the Rx FE is turned on for a short period of time, which saves power. However, with small $K$, it is more probable that all the links in $\mathcal{A}_n$ are blocked.
For a given listening set, the probability of blocking, denoted by $P_B$, can be expressed as follows:
\begin{equation}
    P_B\approx \prod\limits_{k \in \mathcal{A}_n} \mathbb{P}(\gamma_{nk} < \gamma_{\rm min}),
    \label{eq:Pblock}
\end{equation}
where $\gamma_{\rm min}$ is the minimum SNR required for signal detection or decoding. 
This leads to an interesting trade-off. Listening to a large number of cells ensures that the UE has a higher probability of having a usable radio link. But this will require larger wake periods during the DRX cycle, leading to greater power draw. %On the other hand, listening to fewer cells can reduce the monitoring instance interval, $T_{\rm SS}$, but increases $P_B$ and requires more frequent beam sweeping. We explore this trade off in the context of links with blockages in this sequel.

We use the following simple model to quantify
the trade-off.  We assume that
the time to monitor $K$ cells is
$T_{\rm SS}=K T_{\rm SS,0}$, where 
$T_{\rm SS,0}$ is the time to monitor each cell.
If at least one link in the monitoring interval
is not blocked, the UE will spend
a fraction $KT_{\rm SS,0}/T_{\rm SS,per}$ of
the DRX cycle monitoring the link.  
If all links are blocked, we assume that the UE
must stay awake for at least one entire
DRX cycle to perform a complete beam search,
and re-establish a reliable connection.
%The power savings at the UE is quantified in terms of the fraction of time it is asleep, denoted by $\beta_{\rm sleep}$.%, which depends on $K$, the cardinality of $\mathcal{A}_n$. Increasing  $K$ decreases the probability of blockage as seen from (\ref{eq:Pblock}). This ensures fewer beam sweeps.
% Over an interval of length $T_{\rm total}$, let $N_{\rm{IA}}$ denote the number of times that the UE has to perform IA. Similarly, let $N_{\rm{NIA}}$ denote the number of times when IA is unnecessary (i.e., when the UE has a usable radio link in its listening sets). 
%On the other hand, assuming $T_{\rm SS,per}=20~{\rm ms}$ \cite{3gpp_RRC, bmanagement}, the UE measures the channel to the $K$ BSs in $\mathcal{A}_n$ over a period of $(K/N) \times T_{\rm{ss}}$. This cost increases with an increase in $K$.
%In case the UE needs to perform IA, it needs to track all the BSs over a considerably longer interval ($\approx 4\times T_{\rm{SS,per}}$). 
%Hence, over a DRX period of $n_t$ measurement cycles of length $T_{\rm SS,per}$, the fraction of time spent sleeping, $\beta_{\rm sleep}$, is given by
Under the above model, the fraction of time
the UE is awake, $\beta_{\rm awake}$,
and the corresponding time the UE is asleep
$\beta_{\rm sleep}$ can be bounded as,
\begin{align}
\beta_{\rm awake} & \geq (1 - P_B) 
\frac{KT_{\rm SS,0}}{T_{\rm SS,per}} + P_B, \label{eq6}\\
\beta_{\rm sleep} & = 1- \beta_{\rm awake},
\label{eq7}
\end{align}
where the inequality in (\ref{eq6}) results from the fact that the UE might have to be awake for more that $T_{\rm SS,per}$ on link failure. %When $L=1$, the inequality in (\ref{eq6}) is replaced by equality. 
Equations (\ref{eq6})-(\ref{eq7}) shows the trade-off between a the listening set size and power savings in the DRX cycle. This is studied through  cellular simulations both at $28$ and $140$ GHz in the following section.
\begin{algorithm}[b!]
 \KwData{$\{\gamma_{nk}: k=1,\cdots,N; n=1, 2, \cdots\}$}
 $\mathbf{Input:}$ $K,\gamma_{\rm{min}}$
 
$\mathbf{Initialization}$
\begin{itemize}
    \item Listen to all cells and choose the one with the highest SNR as the serving cell. 
     \item Choosing listening set $\mathcal{A}_n$: In addition to the serving cell, uniformly select a subset of $K-1$ cells from the remaining $N-1$ cells.
\end{itemize}

$\mathbf{Operation}$ ~$(n\geq 1)$
\begin{itemize}
    \item Measure $\gamma_{nk}$ for $k\in \mathcal{A}_n$. Let $\gamma_{n,0}$ denote the SNR of the serving cell.\\
  \eIf{$\gamma_{n,0} > \gamma_{\rm min }$} {
   
    $\gamma_{n+1,0} =\gamma_{n,0}$ $\rightarrow$ \textit{Stick to current BS}
  } {
    \eIf { $\underset{k\in \mathcal{A}_n}{\max} \gamma_{nk} > \gamma_{\rm min}$}{$\gamma_{n+1,0} = \underset{k\in{\mathcal A}_n }{\max}~ \gamma_{nk}$ $\rightarrow$ Change serving cell } {Trigger exhaustive beam sweep: Go back to initialization }

  }
\end{itemize}
 \caption{Proposed Algorithm}
 \label{Rmabalgo}
\end{algorithm}
\section{DRX Measurements and Power Savings}
\label{sec:Algo}
To address the problem of saving power by optimizing the DRX cycle under blockage limited access, in this section we propose a simple algorithm. In Algorithm \ref{Rmabalgo}, for a given $K$, the UE selects the best $K$ cells out of all the $N$ cells before going into DRX mode, i.e., at time $n=0$.  Before going into the DRX cycle, based on the $N$ measurements, the UE selects $K$ BS (the listening set $\mathcal{A}_n$) to monitor during the DRX cycle including the primary associated link. In the DRX cycle the link quality of all the BSs in $\mathcal{A}_n$ are measured. 

Let the SNR of the serving BS at the $n$-th measurement period be $\gamma_{n,0}$. If $\gamma_{n,0} \geq \gamma_{\rm{min}} $, the UE does not change the serving BS. Otherwise, if $\gamma_{n,0} < \gamma_{\rm{min}} $, the UE changes its serving BS and chooses the best BS in $\mathcal{A}_n$ such that 
\beq
\gamma_{n+1,0} = \underset{k\in{\mathcal A}_n }{\max}~ \gamma_{nk}.
\eeq
Since the UE will already have synchronization information from the BSs contained in $\mathcal{A}_n$, it will only need to do a random access to change the serving BS. However, when all the links in $\mathcal{A}_n$ are blocked, i.e., \beq 
\underset{k\in{\mathcal A}_n }{\max} ~ \gamma_{nk} < \gamma_{\rm{min}},
\eeq 
the UE has to go through the beam sweep procedure. The beam sweep procedure for beamformed systems can span over several SS periods. The time taken by beam sweep can hence be given as $T_{\rm BSW} = L\times T_{\rm SS,per}$, where $L\geq 1$. The algorithm is summarized in Algorithm~\ref{Rmabalgo}. 
% which involves tracking \emph{all} the BSs and choosing the best one as serving BS. It is to be noted that the IA procedure is much more time consuming than the relatively simple random access, since the UE has to monitor all the possible BSs and has to be awake for a longer period of time, resulting in dissipation of extra power and also in increased latency. 

\section{Numerical Results}
\label{sec:numerical_results}

\subsection{Simulation Setup}
We generate $100$ channel trajectories at $28{~\rm GHz}$ and $140{~\rm GHz}$ using end-to-end simulations that are mainly based on the 3GPP channel model in \cite{3gpp_model}\footnote{While the specifications of the 3GPP channel model are valid up to $100{\rm GHz}$, we continue to use it even for $140{\rm GHz}$ due to the absence of standardized channel models for the spectrum above 100 GHz}. The UE is located at $(0,0,1.8)$ ${\rm m}$. At $28~{\rm GHz}$, we consider 8 antenna elements at the UE, while at $140~{\rm GHz}$, 64 antenna elements are assumed. For simplicity, we do not consider the impact of array geometry and the associated latency and the power consumption involved in finding the best receive direction. Instead, we assume eigen beamforming to calculate the beamforming gain in (\ref{eq:snr}) (i.e., $\nbw_k$ and $\nbf_k$ are equal to the eigenvector corresponding to the largest eigenvalue of $\nbH^{(n,k)}\nbH^{(n,k)^H}$ and $\nbH^{(n,k)^H}\nbH^{(n,k)}$, respectively). Hence, in effect, we assume that the UE and the BS can always `point' towards each other and determine the strongest direction of the received signal. The challenges posed by array geometry is left for future work.

\begin{table}[b]
\centering
\begin{tabular}{|c|c|c|} 
\hline
\textbf{Parameters}  & \textbf{28 GHz} & \textbf{140 GHz}                   \\ 
\hline
Scenario    & \multicolumn{2}{c|}{UMi}        \\ 
\hline
$M_u$ &  $8$  & $64$                 \\ 
\hline
$M_b$ & $64$  & $256$                      \\ 
\hline
BS height   & \multicolumn{2}{c|}{$10~{\rm m}$}         \\ 
\hline
 $N$          & \multicolumn{2}{c|}{$9$}        \\ 

\hline
UE height   & \multicolumn{2}{c|}{$1.8{\rm m}$}        \\ 
\hline
Bandwidth   & \multicolumn{2}{c|}{$400~{\rm MHz}$}      \\ 
\hline
Sampling interval & \multicolumn{2}{c|}{$20~{\rm ms}$}         \\ 
\hline
Temperature          & \multicolumn{2}{c|}{$298~{\rm K}$}        \\ 

\hline
Cell radius, $r$   & \multicolumn{2}{c|}{$100~{\rm m}$}        \\ 
\hline
Blocker density, $\lambda_b$     & \multicolumn{2}{c|}{$0.01~{\rm m}^{-2}$} \\ 
\hline
Blocker height       & \multicolumn{2}{c|}{$1.4~{\rm m}$~(Vehicular), $1.7~{\rm m}$~(Human)}    \\ 
\hline
Blocker width        & \multicolumn{2}{c|}{$4.8~{\rm m}$~(Vehicular), $0.3~{\rm m}$~(Human)}    \\ 
\hline
Blocker speed        & \multicolumn{2}{c|}{\begin{tabular}{@{}c@{}} $0$-$28~{\rm m/s}~(0$-$100~{\rm km/h}) [\mbox{Vehicular}]$ \\ $0$-$1~{\rm m/s}~(0$-$3~{\rm km/h}) [\mbox{Human}]$\end{tabular}}   \\ 
\hline
Transmitted Power    & \multicolumn{2}{c|}{$23{\rm dBm}$}       \\
\hline
\end{tabular}
\caption{Values of different parameters for the generation of channel trajectories }
\label{tabel_params}
\end{table}
 
%and is assumed to have two antenna arrays - one each on its top and side edges, similar to the Edge model considered in \cite{qualcomm}. Only one of the two arrays is assumed to be active in any paging cycle, depending on the orientation of the UE (i.e., if the UE is in portrait (landscape) mode, then the top (side) array is active). At $28~{\rm GHz}$, we consider the array geometry at the UE to be a $4\times2$ Uniform Rectangular Array (URAs), while at $140~{\rm GHz}$, a $64\times1$ Uniform Linear Array (ULAs) is assumed. 

%It is assumed that the channel matrix is known at the UE and beamforming gain is calculated using eigenvector associated to the maximum eigenvalue.

We consider a cell radius, ($r$), of $100~{\rm m}$ and deploy 9 BSs of height $10{\rm m}$ in the $xy$-plane on a square grid of dimensions $\rm{200~m\times200~m}$. The BSs can either be in line-of-sight or non-line-of-sight. At $28~{\rm GHz}$ and $140~{\rm GHz}$, we consider the number of antennas at the BS to be 64 and 256 respectively. The UE is dropped randomly with in the grid for a given channel trajectory and remains stationary.

We distinguish between two kinds of blockers - human and vehicular, which are assumed to be present with equal probability. The blockage loss is modeled using the 3GPP blockage model B \cite{3gpp_model}, which is based on double knife-edge diffraction, where each blocker is modeled as rectangular screen in the vertical plane. The blocker dimensions are chosen according to 3GPP recommendations \cite{3gpp_model}, with the height and width of a human (vehicular) blocker set to $1.7~{\rm m}$ ($1.4~{\rm m}$) and $0.3~{\rm m}$ ($4.8~{\rm m}$), respectively. For a given channel trajectory, let $\nbx_j(n) \in \nbbR^2$ denote the projection of the centroid of the $j$-th blocker onto the $xy$-plane during the $n$-th monitoring instance. At $n=0$, the collection of obstacle locations, $\{\nbx_j(0)\}$, are distributed according to a Poisson point process of intensity $\lambda_b = 0.01~\rm{m^{-2}}$ \cite{ishblockage} over a circle of radius $200~\rm{m}$ centered at the UE. For $n\geq 1$, $\nbx_j(n)$ evolves in a Markovian manner as follows:
\beq 
 \nbx_j(n) = \nbx_j(n-1) + \dot{\nbx}_j(n)\Delta,
\eeq
where $\dot{\nbx}_j(n) \in \nbbR^2$ denotes the velocity along the $xy$-plane of the $j$-th obstacle during the $n$-th monitoring instance and $\Delta=20~{\rm ms}$ is the sampling period (SSB periodicity \cite{bmanagement, 3gpp_RRC}).

The initial blocker velocities, $\{\dot{\nbx}_j(0)\}$, are drawn independently and uniformly over $[0,1] ~\rm{m/s}$ for a human blocker and over $[0,28]~{\rm m/s}$ for vehicular blockers\footnote{These velocity ranges for human and vehicular blockers are based on 3GPP recommendations, as well \cite{3gpp_model}.}. For $n\geq 1$, $\dot{\nbx}_j(n)$ evolves in the following manner:
\beq
 \dot{\nbx}_j(n) = \dot{\nbx}_j(n-1) +\nbw(n)
\eeq
where $\{\nbw(n):n\geq 1\}$ is a sequence of i.i.d zero-mean Gaussian random vectors with identity covariance matrix. For simplicity, we assume $\nbw_k$ and $\nbf_k$ to be equal to the left and right singular vectors corresponding to the largest singular value of $\nbH^{(n,k)}$, respectively.

 % and similar model is also mentioned in \cite{metis_model}
     
%\begin{figure}[!t]
 %   \centering
  %  \includegraphics[width=70mm]{blkageforpaper.png}
  %  \caption{Example of blockage events in a channel trajectory at 28 GHz}
  %  \label{fig:blockfig}
%\end{figure}
%A sample channel trajectory at 28 GHz is shown in Fig. \ref{fig:blockfig}, where the temporal correlation in the SNR is especially evident during blocking events. Each trajectory consists of $10^{4}$ SNR values.
The list of all the parameters used for generating the channel trajectories are provided in Table \ref{tabel_params}.    

%, sampled at $5~{\rm ms}$, which is one of the suggested values for the SS-burst period in 3GPP-NR \cite{bmanagement}.

\subsection{Simulation Results}
%We compare the performance of the Algorithm 1 with that of UCB1 algorithm which assumes i.i.d rewards. The SNR is discretized by rounding to the nearest dB. The discretized SNR is then converted to linear scale and is the input to both the algorithms. 

 %The time-averaged regret for UCB1 and restless multi-armed bandit (RMAB) is plotted in in Fig. \ref{fig:performanceregret} where it can be observed that Algorithm 1 achieves sub-linear regret since $r(n)/n$ decays to zero. This ensures that the UE learns the strongest BS over time and keeps listening to it often. On the other hand, UCB1 only achieves linear regret since $r(n)/n$ converges to a non-zero quantity. This serves to validate our restless MAB formulation and demonstrates the importance of considering the temporal correlation in the channel qualities while devising energy saving policies for choosing the listening set.

%where the , the blockage probability ($P_B$) and fractional UE sleeping time, $T_{\rm sleep}$, are quantities of interest as a function of $K$. Based on the results, we identify the smallest value $K$ that guarantees a usable radio link with a probability greater than $99\%$.

The performance of Algorithm~\ref{Rmabalgo} is evaluated using the SNR from the generated channel trajectories. The quantities of interest, as a function of $K$, are the blocking probability, $P_B~(\gamma_{\rm min}=-6.5 \mbox{ dB})$ and the fractional UE sleeping time, $\beta_{\rm sleep}$.

\begin{figure}[!t]
    \centering
    \includegraphics[width=0.4\textwidth]{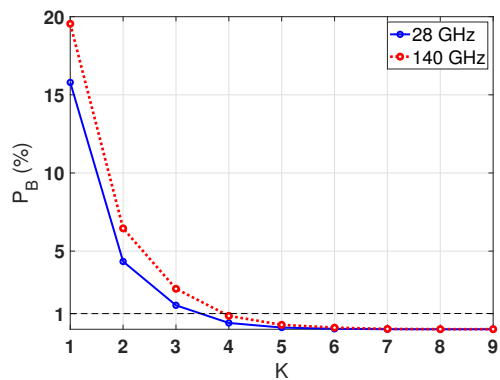}
    \caption{Variation of $P_B$ as a function of $K$. The dashed horizontal line corresponds to $P_B=1\%$.}
    \label{fig:probability_block}
\end{figure}

% $P_-$ is minimum if the UE tracks all the BSs as mentioned in Sec. \ref{sec:problem_formulation} and is maximum when the UE tracks only one BS i.e. $K=1$. For 28 GHz(140 GHz), the minimum $P_-$ corresponds to no blockage while the maximum is $15.7\% (19.5\%)$.
Fig.~\ref{fig:probability_block} captures the variation in $P_B$ as a function of $K$. We observe that $K=4$ is sufficient to guarantee a usable radio link with a probability greater than $99\%$. 

% Why do we even care about these cdfs? All that we are interested in is the probability that $\gamma_o > \gamma_{\rm min}$ and Fig. 5 adequately answers that question, does it not?

% The cdf of $\gamma_o$ is plotted in Fig.~\ref{fig:performanceregret}. While listening to fewer BSs comes at the expense of a smaller value of $\gamma_o$, the impact of this is minimal since the magnitude of $\gamma_o$ is not particularly important for DRX mode operation as long as it is above $\gamma_{\rm min}$. Thus, there is a diminishing return to listening to more BSs. We observe that in terms

%because the current listening set will not always have the best BS until it tracks them all. On the other hand, tracking all BSs has a power cost associated to it which is higher than tracking fewer BSs.

\begin{figure}[!t]
    \centering
    \includegraphics[width=0.4\textwidth]{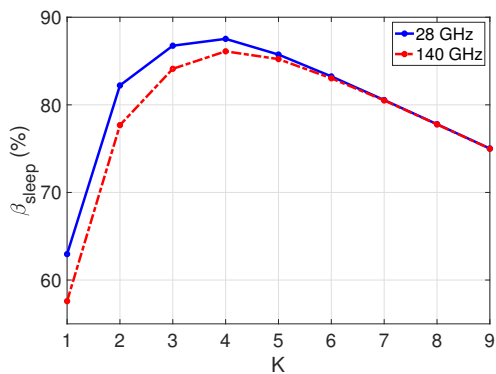}
    \caption{Fractional UE sleep time, $\beta_{\rm sleep}$, as a function of $K$.}
   \label{fig:sleep_time}
\end{figure}

%\begin{figure} [!t]
%\centering
%\begin{subfigure} {0.23\textwidth}
%\centering
%	\includegraphics[scale=0.23]{figs/cdf28.eps}
%	\caption{28 GHz.}\label{fig:reg_28GHz}
%\end{subfigure}
%\quad
%\begin{subfigure} {0.23\textwidth}
%\centering
%	\includegraphics[scale=0.23] {figs/140cdf.eps} \caption{140 GHz.}\label{fig:reg)140GHz}
%\end{subfigure}
%\caption{The CDF of the SNR on the serving link determined by Algorithm~\ref{Rmabalgo}.}
%\label{fig:performanceregret}
%\end{figure}

% \begin{figure}[!t]
%  \centering
% \begin{subfigure}{0.2\textwidth}
% \centering
%   \includegraphics[scale=0.25]{figs/cdf28.eps}
% \caption{28 GHz}
% \label{fig:reg_28GHz}
% \end{subfigure}
% \qquad\qquad
% \begin{subfigure}{0.2\textwidth}
% \centering
%     \includegraphics[scale=0.25]{figs/140cdf.eps}
% \caption{140 GHz}
% \label{fig:reg)140GHz}
% \end{subfigure}
% \caption{The CDF of the SNR on the serving link as determined by Algorithm~\ref{Rmabalgo}.}
% \label{fig:performanceregret}
% \end{figure}
Fig.~\ref{fig:sleep_time} plots $\beta_{\rm sleep}$ as a function of $K$. While it seems intuitive to expect that $\beta_{\rm sleep}$ would be the highest for $K=1$, since $T_{\rm SS}$ has the smallest value in this case, this is not true in reality since the probability that the UE has to perform exhaustive beam sweep (equal to $P_B$) is also high. At the other extreme when $K=9$, even though the UE does not have to perform beam sweep often, $\beta_{\rm sleep}$ is still not the highest due to the large value of $T_{\rm SS}$. From Fig.~\ref{fig:sleep_time}, we observe that $\beta_{\rm sleep}$ is highest ($> 85\%$ for 28 GHz and 140 GHz) when $K=4$. Thus, from Figs.~\ref{fig:probability_block} and \ref{fig:sleep_time}, we infer that $K=4$ is sufficient to guarantee a usable radio link at least $99\%$ of the time while saving the most amount of power.

\section{Conclusion}
\label{sec:concl}
A central challenge for realizing mobile and
handheld devices in the mmWave
and THz frequencies is the high power consumption
of the RFFE.
Aggressive use of DRX modes offers the possibility
of significant power savings, assuming
traffic is bursty.  However, mmWave and THz
channel quality can be intermittent due to blockage.
frequent channel monitoring in DRX
mode thereby reducing the power savings.
In this work, we have presented preliminary 
estimates of RFFE power consumption for DRX
measurements.  A simple algorithm to minimize
the number of cells to monitor has also been proposed
and simulated at both 28 and 140~GHz.
Our simulations indicate that, using correct
predictions, a small number of cells can be tracked,
while maintaining high levels of reliability.

\section*{Acknowledgements}
This work was supported by the 
National Science Foundation under Grants
1302336, 1564142, and 1547332, 
NIST, SRC and the industrial
affiliates of NYU WIRELESS.

\balance
\bibliographystyle{IEEEtran}
\bibliography{IEEEabrv,mybib}

\end{document}

%% file: notation_new.tex
\def\nbf{{\mathbf{f}}}

\def\nbw{{\mathbf{w}}}
\def\nbx{{\mathbf{x}}}

\def\nb0{{\mathbf{0}}}
\def\nb1{{\mathbf{1}}}

\def\nbH{{\mathbf{H}}}

\def\nbbC{{\mathbb{C}}}

\def\nbbR{{\mathbb{R}}}